# Optical repumping of resonantly excited quantum emitters in hexagonal boron nitride

Simon J. U. White,[1,*] Ngoc My Hanh Duong,[1,*] Alexander S. Solntsev,[1] Je-Hyung Kim[3], Mehran Kianinia[1^], and Igor Aharonovich[1,2,^]

[1]*School of Mathematical and physical sciences, University of Technology Sydney, Ultimo, NSW 2007 Australia*
[2]*ARC Centre of Excellence for Transformative Meta-Optical Systems, University of Technology Sydney, Ultimo, New South Wales, Australia*
[3]*School of Nature Science, Department of Physics, Ulsan National Institute of Science and Technology (UNIST), Ulsan 44919, Republic of Korea*

[^]Mehran.Kianinia@uts.edu.au  Igor.Aharonovich@uts.edu.au
*Authors contributed equally

*Abstract:* Resonant excitation of solid-state quantum emitters enables coherent control of quantum states and generation of coherent single photons, which are required for scalable quantum photonics applications. However, these systems can often decay to one or more intermediate dark states or spectrally jump, resulting in the lack of photons on resonance. Here, we present an optical co-excitation scheme which uses a weak non-resonant laser to reduce transitions to a dark state and amplify the photoluminescence from quantum emitters in hexagonal boron nitride (hBN). Utilizing a two-laser repumping scheme, we achieve optically stable resonance fluorescence of hBN emitters and an overall increase of ON time by an order of magnitude compared to only resonant excitation. Our results are important for the deployment of atom-like defects in hBN as reliable building blocks for quantum photonic applications.

Solid-state single photon emitters that possess narrow emission linewidths are promising for scalable quantum photonic application[1-3]. Specifically, defects in diamond[4], silicon carbide[5] or rare-earth ions in solid-state hosts[6,7] are attracting considerable attention due to their potential use in spin-photon interface architectures[8]. In recent years, single defects in atomically thin materials such as transition metal di-chalcogenides or hexagonal boron nitride (hBN) have been identified[9-11]. The hBN quantum emitters possess a high brightness at room temperature and often exhibit a high Debye-Waller factor that maximizes the emission in its zero phonon line (ZPL)[12-15]. Significant progress has been achieved in performing spectroscopic studies of hBN quantum emitters[15-20], and modulating their emission wavelengths using strain or electric field[21-23].

However, under resonant excitation, the hBN emitters exhibit blinking and undergo a transition to a metastable state, that often results in a non-radiative decay[24-26]. The blinking behaviour of hBN quantum emitters prohibits efficient resonant excitation and coherent manipulation, that is needed for scalable quantum applications. Similar blinking behaviour was also reported for quantum dots[27,28] and some of the color centres in diamond[29,30]. Indeed, a lot of attention has been devoted to improve the stability of the main optical transitions under resonant excitation. Various methods include dynamic electric field modulation or active feedback that stabilises the random spectral diffusion of the emitter[28,31], or optical repumping of the optical transition of the selected emitter to avoid relaxation to the dark state[29,30].

In this work, we employ a weak non-resonant laser at a wavelength between 500 nm and 532 nm, in addition to the resonant laser that drives the system coherently to stabilise the optical transition of single defects in hBN. The secondary laser acts as an additional excitation pathway and re-initializes the system into its bright state. The two laser repumping scheme enables the observation of bright resonant fluorescence with Fourier Transform limited photons emitted from the hBN. We provide an in-depth analysis of the photodynamics of the system, and discuss its applicability for an improved coherence and future photon indistinguishability measurements.

The optical measurements are performed by using a home-built confocal microscope with a Michelson interferometer at the collection, as shown in Fig. 1(a). The investigated sample consists of hBN flakes (Graphene supermarkets) drop cast on a silicon substrate with native oxide, annealed at 850°C for 30 minutes to activate the emitters. The sample is placed inside a closed-loop liquid helium flow cryostat (Attocube Attodry800) and cooled down to ~4 K. A Ti:sapphire laser (Msquared) with a linewidth of ~50 kHz was directed into the objective as an excitation source. In addition, a 532 nm (Laser quantum, GEM) or a tunable laser (NKTphotonics, SuperK Fianium,) was aligned to the excitation path for co-excitation with Ti:sapphire (Fig. 1(a)). The emission was collected either into a spectrometer or avalanche photodiodes (Excelitas SPCM-AQRH) in a Hanbury Brown and Twiss configuration.

We start with off resonant excitation of the system. To investigate the effect of a weak non-resonant laser, we characterize the ZPL of the emitters with incoherent excitation at 715 nm as well as a 532 nm weak non-resonant laser. Figure 1(b) shows the photoluminescence (PL) spectra of a selected hBN emitter with a ZPL at ~ 774 nm, under two excitation conditions: (i) 715 nm laser (red trace, 650 µW), and (ii) co-excitation of 715 nm + 532 nm lasers (green trace, 650 µW, and 60 µW respectively). The PL emission of the emitter displays a narrow peak with FWHM of ~ 0.11 nm and a ZPL at ~ 774 nm driving from a Lorentzian fit to the data. Under excitation with 532 nm (60 µW), there was no detectable fluorescence from this emitter. However, adding the 532 nm laser to the 715 nm laser results in an increase of the emission intensity by nearly two-fold[18,32]. The results below are resented for this selected emitter, but we observed a similar behaviour from other probed emitters. Note that due to the operation of the Ti:sapphire laser, we limited the studies of hBN defects with ZPLs above 700 nm.

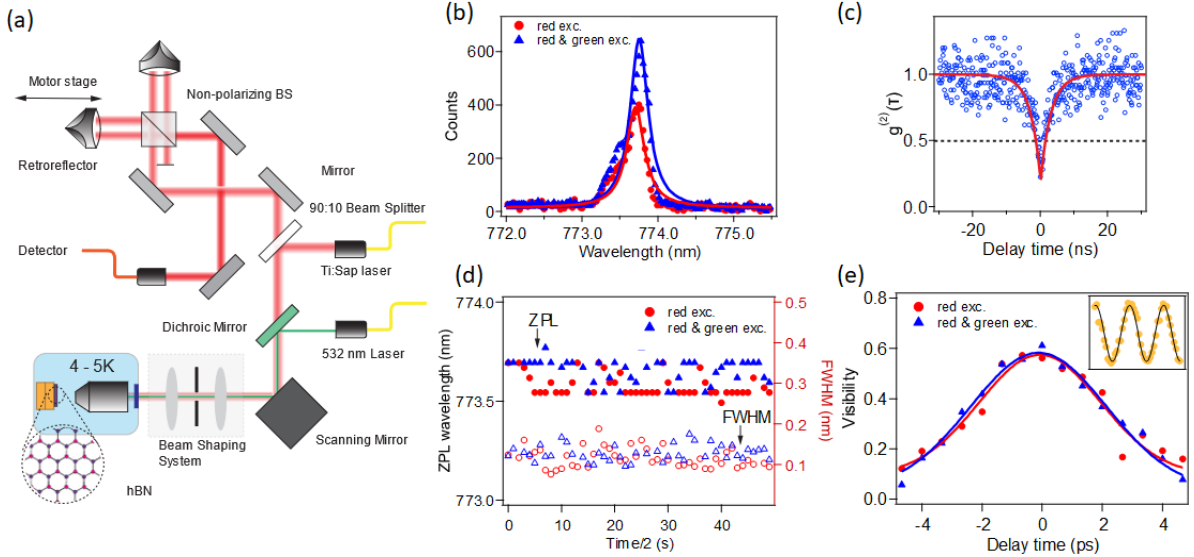

*Figure 1: Cryogenic photoluminescence measurement. (a) Schematic of the PL setup with two CW excitation sources (532 nm, 715 nm) and a Michelson interferometer at the collection. (b) PL spectra of the emitter with 715 nm excitation (red) and repumping with 532 nm (green). The solid lines are Lorentzian fit to the ZPL yielding the linewidth of 135 GHz. (c) Second-order correlation measurement under the repumping condition. (d) PL spectroscopy statistic of the emitter with 715 nm excitation only (red) and co-excitation of 715 nm with 532 nm (blue) over a period of 50s. (e) Coherence time measurement of red excitation (red circles) and the co-excitation with a 532 nm (blue triangles). Inset: an example of intensity fluctuation at near-zero arm length different positions)*

Second-order correlation measurement under co-excitation (Fig 1.(c)) indicates single-photon antibunching behavior with $g^2(0) = 0.23$ without any background correction. In this work, we focus on the effect of the co-excitation scheme on the coherent properties of the emitters and its blinking dynamics. Under excitation with 715 nm (Fig 1.(d), red trace) the typical spectral diffusion in hBN emitters is observed[ref]. The spectral diffusion range is around 0.25 nm and full width at half maximum (FWHM) varies within ~ 22 GHz. Each spectrum was collected for 2 seconds. Under co-excitation with the green laser there is no noticeable difference and the emission linewidth and the ZPL exhibit a similar behavior, as shown by the blue traces, Fig 1(d).

To further understand the effect of the two laser co-excitation, we measured the coherence time of the emitter under co-excitation schemes using the Michelson interferometer depicted in Fig. 1(a). For this measurement, the piezo was scanned over two interference fringes with an integration time of 200 ms on each point (Fig 1(e) inset). Figure 1(e) is the plot of the coherence visibility, i.e. the interferogram for various path differences acquired for excitation with 715 nm laser (red trace) or co-excitation (blue trace). The interferogram is fitted with a Gaussian model with the center position set at the equal arm length to extract the coherence times. The coherence time of 6.8 ± 0.4 ps corresponds to red excitation and values of 6.4 ± 0.4 ps was obtained under co-excitation, corresponding to linewidths of ~ 147 and 163 GHz, respectively. Although there is a slight decrease in the coherence time, it is well within the uncertainty of the measurement, albeit lower than the typical coherence values from defects in hBN[33]. Importantly, the comparable values indicate that the ultrafast spectral diffusion, i.e. the inhomogeneous broadening of the natural linewidth, remains on the same magnitude.

We then performed off-resonant photon correlation measurements to get better insights into the co-excitation effect. Figure 2(a) shows the photon correlation measurement with the excitation at the wavelength of 715 nm with the power of 650 µW (red circles), and co-excitation with the wavelength of 500 nm at 60 µW power (blue circles). As expected, the second-order autocorrelation function exhibits antibunching at a short delay time (ns range) and bunching at a longer delay time (µs range) corresponding to long-lived dark states. The best fit to the data was acquired using three long-lived metastable states. The fitting yields the values of $\tau_1$ = 41±0.65 µs (27±0.60 µs), $\tau_2$ = 0.35±0.08 ms (0.26±0.03 ms) and $\tau_3$ = 4.5±1.8 ms (1.3±0.7 ms) for the single 715 nm laser excitation (co-excitation), respectively. The amplitude of the bunching component of the blue trace (co-excitation) decreases, showing an agreement with the enhancement of PL intensity of the emitter upon adding the second laser (Fig. 1(b)). This measurement confirms that in the case of the excitation only with one 715 nm laser, the system spends more time in dark metastable states.

Furthermore, photon statistics show the existence of a very fast blinking in the system (in the order of ~10 ms). Figure 2(b) shows the PL intensity time trace binned every 10 ms along with the corresponding histograms on the right-hand side. In the case of 715 nm excitation, the system undergoes blinking events at the ms timescale which is suppressed by adding the green co-excitation, as confirmed by the reduced bunching and the improved histogram shown in Fig. 2(b).

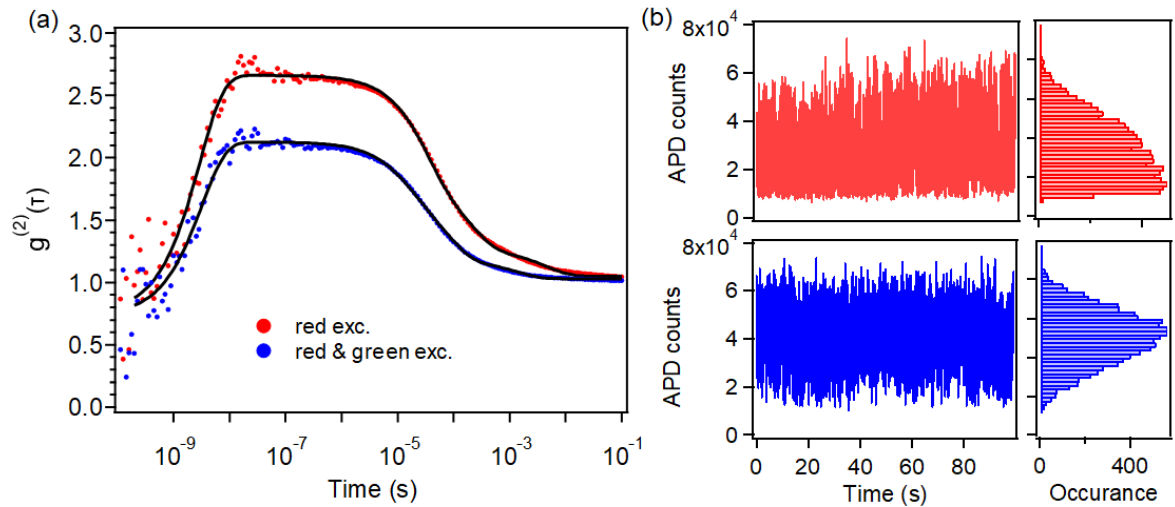

*Figure 2: (a) Photon correlations showing noticeable intermediate states without (red dots) and with co-excitation (blue dots) respectively. The black lines are the fitting of the correlation data with a second order correlation function corresponding to a four-level system. (b) Time trace of APD counts (bin width of 10 ms) and the corresponding histogram showing the reduction of blinking (number of bins: 50 bins, integration: 1350 s), without (top figures) and with (bottom figures) the co-excitation scheme*

Leveraging the performance of the emitter under co-excitation, we now attempt coherent excitation of the emitters. We employed the co-excitation scheme to resonantly excite the emitter, by tuning the Ti:sapphire tunable laser to the ZPL wavelength. In this scheme the co-excitation is achieved by the combination of Ti:sapphire and a green laser emitting at 500 nm at 1 μW. To minimize laser scattering, we spectrally filtered the resonant laser and only collected phonon sideband using a long pass filter at 780 nm. Figure 3(a) shows the PL intensity, from the emitter upon, resonant co-excitation with 30 s pulses of the repumping laser, over a period of 5 minutes (blue trace, offset by +1 kHz), compared to the same sequence with the lasers focussed off the emitter (black trace). The bottom panel of figure 3(a) shows a schematic of the laser sequence. Interestingly, only when a green laser is added to the resonant pump, the emitter can be resonantly excited and detected repeatedly. In fact, under resonant excitation without any repumping, the probability of exciting resonantly is very low due to spectral diffusion (spectral jumps due to charge fluctuations around the emitter) or blinking (entering long-lived dark states).

The analysis of photoluminescence intensity upon resonant excitation with or without repumping is depicted in Fig. 3(b). To define the ON- and OFF-resonance times (denoted as $\tau_{on}$ and $\tau_{off}$, respectively), we use a threshold values of 1.5 kHz and 2.5 kHz (black dashed lines) for the resonant excitation only and the co-excitation scheme, respectively. In case of only resonant excitation, only 1% of the time ($\tau_{on}/\tau_{off} = 0.01$) the emitter is excited while adding the green repumping, the probability increases to more than 15% ($\tau_{on}/\tau_{off} = 0.17$), which is an over order of magnitude improvement. Although the probability of the system being "ON" is dramatically higher under the co-excitation, the system stays on resonance for a relatively short period of 0.1 seconds. Nevertheless, this is sufficient to generate over $10^5$ photons before the next spectral jump.

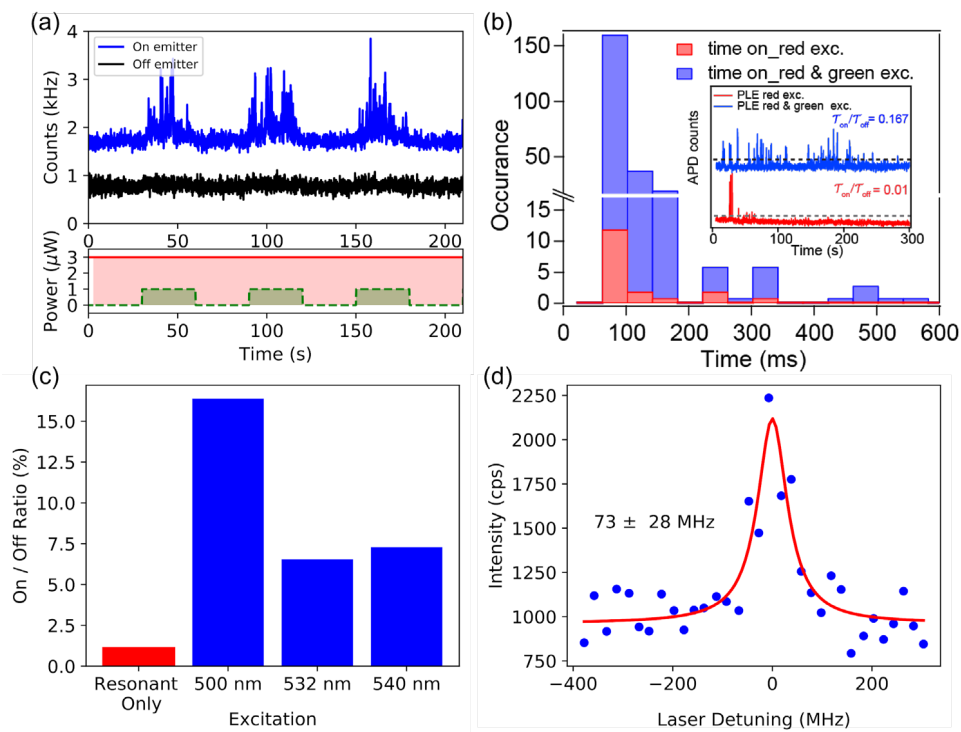

*Figure 3: Resonant excitation of the emitter. (a) Resonant fluorescence intensity trace of optical transition with and without an additional non-resonant repumping laser. The resultant switching PL corresponds to 30 s on and off durations of the 500 nm laser focussed on the*

*emitter (blue trace, offset by 1 kHz) and away from the emitter for reference (black trace). (b) Histogram of ON-Resonant time with resonant excitation 1 μW (red bar) and with co-excitation with 500 nm, 1 μW (blue bar). Inset is PL intensity traces of resonant excitation with 1 μW of 500 nm laser (blue) and without (red) repumping. (c) Effect of the wavelength of the co-exciting laser, showing the 500 nm wavelength is the most efficient. (d) Resonant PLE with 500 nm repumping laser showing near lifetime limited linewidth of 73 MHz.*

We have also compared other co-excitation wavelengths, including the most commonly used 532 nm. We found a strong dependence on the repumping laser, and the 500 nm (~ 2.48 eV) works significantly better than the lower energy lasers at 532 nm or 540 nm. The measured value of ~ 2.48 eV can correspond to either a charge transfer threshold of the particular color centre, or would be sufficiently high to charge the nearby traps that would result in a sufficient free carrier density to be trapped by the quantum emitter[19]. While we cannot distinguish between the two scenarios, the second scenario is more likely since co-excitation only increases the emitter brightness, as was shown for two off resonant excitation cases (Figure 1). Note that lower co-excitation wavelength (e.g. 480 nm), also showed a reduced repumping efficiencies compared to the 500 nm one (not shown).

More importantly, however, each initialisation of the co-excitation cycle results in a flux of photons, while the resonant laser frequency remains constant. This indicates that the emitters' original ZPL position remains the same, without any significant spectral drift. This is vital for the future generation of indistinguishable photons from the hBN defects[26]. In this regard, we were able to extract the linewidth of the emitter by scanning the resonant laser over the ZPL, while the green laser is on, we were able to measure the temporal linewidth of the emitter, as shown in figure 3c. The obtained linewidth is 73 ± 28 MHz, corresponding to the Fourier Transform limit of single-photon emitters in hBN with typical lifetimes of a few ns. Under the green co-excitation, the blinking events (in ms range) are reduced significantly and the probability of resonant excitation is significantly increased.

To summarize, we have demonstrated an effective co-excitation scheme for the resonant excitation of hBN quantum emitters. We showed that by applying the two-laser repumping scheme, we can recover emitters from the dark state, resulting in increased resonant photoluminescence. The probability of emission, in this case, is increased by a factor of 15. Importantly, this approach can enhance the brightness of single-photon emitters and be effectively use to increase the *on-resonance* time of the emitter. Given the progress with electrical field modulation of quantum emitters in hBN, future studies may achieve stabilisation of coherent emission electrically, thus paving the way for on-chip devices. Our work serves as an important step towards the utilization of hBN quantum emitters for quantum information applications.

**Acknowledgements**
We thank Dr Carlo Bradac for useful discussions. The authors acknowledge the Australian Research Council for (DP180100077) and the Asian Office of Aerospace R&D (FA2386-20-1-4014) for the financial support.